\documentclass[journal]{IEEEtran}
\ifCLASSINFOpdf
\else
\fi
\usepackage{subfigure}
\usepackage{graphicx}% Include figure files
\usepackage{dcolumn}% Align table columns on decimal point
\usepackage{bm}% bold math
\usepackage{epsfig}
\usepackage{epstopdf}

\begin{document}
\title{Hydrogen Concentration in Photovoltaic a-Si:H Annealed at Different Temperatures Measured by Neutron Reflectometry}
\author{A.~J.~Qviller,
        E.~S.~Marstein,
        C.~C.~You,
        H.~Haug,
        J.~R.~P.~Webster,
        B.~Hj\"{o}rvarsson,
        C.~Frommen and
        B.~C.~Hauback% <-this % stops a space
\thanks{A. J. Qviller (corresponding author, e-mail: atlejq@gmail.com), E. S. Marstein, C. C. You, H. Haug, C. Frommen and B. C. Hauback are with Institute for Energy Technology, P.O. Box 40, NO-2027 Kjeller, Norway}% <-this % stops a space
\thanks{J. R. P. Webster is with ISIS, STFC Rutherford Appleton Laboratory, Harwell Oxford, United Kingdom}
\thanks{B. Hj\"{o}rvarsson is with Division for Materials Physics, Department of Physics and Astronomy, Uppsala University, Box 516, SE-751 20 Uppsala, Sweden}
%\thanks{Manuscript received April 30, 2017; revised August 31, 2017.}
}

\markboth{$\copyright$ 2018 IEEE}%
{Qviller \MakeLowercase{\textit{et al.}}: Hydrogen Concentration in Photovoltaic a-Si:H Annealed at Different Temperatures Measured by Neutron Reflectometry}

\maketitle

\begin{abstract}
Amorphous hydrogenated silicon (a-Si:H) is an important material for surface defect passivation of photovoltaic silicon (Si) wafers in order to reduce their recombination losses. The material is however unstable with regards to hydrogen (H) desorption at elevated temperatures, which can be an issue during processing and device manufacturing. In this work we determine the temperature stability of a-Si:H by structural characterization of a-Si:H/Si bilayers with neutron- (NR) and X-ray reflectometry (XRR) combined with photoconductance measurements yielding the minority carrer lifetime. The neutrons are sensitive to light elements such as H, while the X-rays which are insensitive to the H-concentration, provide an independent constraint on the layer structure. It is shown that H-desorption takes place at a temperature of approximately T = $425\,^{\circ}\mathrm{C}$ and that hydrogen content and minority carrier lifetimes have a strongly correlated linear relationship, which can be interpreted as one hydrogen atom passivating one defect. %Both reflectometry techniques have a sub-nanometer resolution and allow non-destructive probing of buried interfaces.
\end{abstract}

\begin{IEEEkeywords}
Solar energy, Silicon, Amorphous materials.
\end{IEEEkeywords}

\section{Introduction}
\IEEEPARstart{O}{ne} of the most important trends in the solar cell industry is the development of solar cells with higher efficiencies. Two of the most promising architectures for obtaining this using wafer-based Si solar cells are based on the use of hydrogenated amorphous Si (a-Si:H): the surface passivation of a-Si:H thin films is essential for the development of both Si heterojunction (HJ) and more recently Si heterojunction back contact (HJBC) solar cells. The latter has enabled Kaneka to manufacture Si solar cells with efficiency above 26\% under one Sun, the current world record \cite{Green}. In addition to the crystalline silicon-based heterojunction structures, a-Si:H is also a successful thin film solar cell technology in its own right, both in single- and multijunction architectures. These technologies have also seen efficiency improvement in recent years. The a-Si:H thin films are commonly fabricated using plasma-enhanced chemical vapor deposition (PECVD) \cite{Wolf} and it is well known that the H incorporated in the material during deposition plays an important role in the passivation of surface-related defects. a-Si:H has one major drawback: it is unstable for use at high temperatures \cite{Li}. This makes it less useful for conventional solar cell architectures which rely on high temperature firing processes. H-concentration in a-Si:H is currently measured by several different techniques \cite{Abou-Ras}, but they generally require thick films in order to get a good signal or have the challenge of limited depth resolution. In order to better understand the role of H in the near surface passivation process, as well the process of thermally induced degradation, we have investigated surface-passivated Si wafers with a-Si:H thin films before and after thermal treatment using a combination of electronic, structural and chemical characterization methods. The determination of minority carrier lifetime was obtained by quasi-steady-state photoconductance (QSSPC) measurements. A combination of neutron reflectivity (NR) and X-ray reflectivity (XRR) have been used for structural and chemical characterization as described by Qviller et al.\cite{Qviller}. Here we report on an extensive investigation on the effect of thermal treatment and we demonstrate a linear relation between the H concentration and the near surface surface passivation.

\section{Experiments}

\subsection{Sample preparation}

The a-Si:H films were deposited on float-zone wafers made of n-type silicon with the $\langle$100$\rangle$ orientation. The wafers were dipped in hydrogen fluoride (HF) (5\%) and rinsed in de-ionized water before deposition. Afterwards, a-Si:H was deposited on each side of the wafer in an Oxford Systems Plasmalab 133 PECVD reactor at $230 \,^{\circ}\mathrm{C}$ by the use of silane gas (SiH$_{4}$). Samples were thereafter annealed in a rapid thermal processing (RTP) system (AccuThermo AW610) for 1 min at steady state. A WTC-120 Photoconductance Lifetime Tester from Sinton Instruments was used to measure the minority carrier lifetime.

\subsection{Reflectivity measurements}

Specular reflectometry is a scattering technique that can be used to reconstruct the scattering length density (SLD) as a function of distance above the substrate ($z$) in thin films. The SLD  for neutrons is defined as

\begin{equation}
SLD = \frac{1}{\phi} \sum\limits_{i} n_{i}b_{i}
\label{SLD}
\end{equation}

where ${\phi}$ is the molecular volume, $n_{i}$ is the number of atoms of element $i$ in this volume and $b_{i}$ is the corresponding coherent scattering length. X-ray SLDs are similarly defined as the volume average of the sum of electron densities for all present elements.

All scattering techniques suffer from the \emph{phase problem} which results in a lack of uniqueness in the determined elemental distribution \cite{XNR2}. To remedy this shortcoming, one can, for example, utilize a combination of neutron and X-ray measurements, in addition to a priori information about the sample, such as layer ordering and approximate thicknesses of the layers. For a-Si:H/Si bilayers exposed to oxygen (O), the neutrons are sensitive to the concentration of H, Si and O and, therefore, provide information on the concentration and gradients in these elements, while the contrast between Si and SiO$_{2}$ is small. The X-ray data can therefore be used as a constraint on the obtained chemical gradients and also be used for determining the total thickness of the a-Si:H and SiO$_{2}$ layers. % and the presence of H is only obtained through changes in the density of Si.

\begin{figure}[!t]
  \centering
    \includegraphics[scale=0.37]{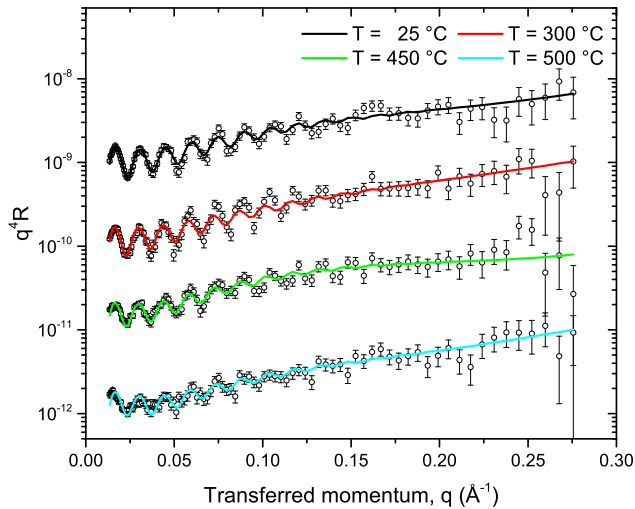}
    \caption{NR reflectivity profiles and model fits for the a-Si:H/Si samples. The graphs are offset from each other by a factor of 0.1 for clarity. Data and fits are footprint corrected.}
    \label{neutrons40}
\end{figure}

\begin{figure}[!t]
    \includegraphics[scale=0.37]{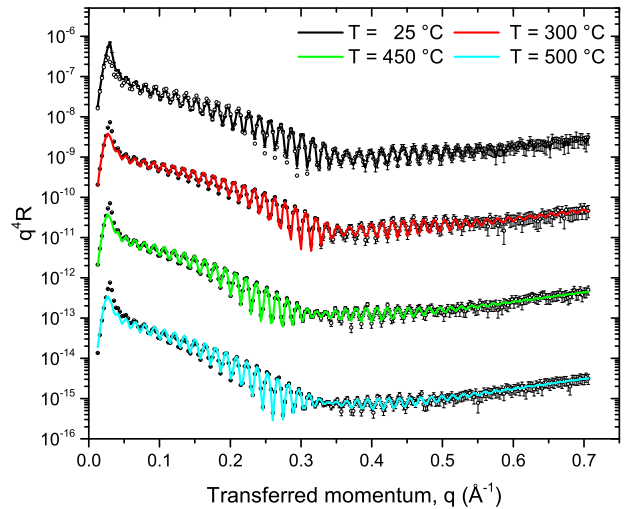}
    \caption{XRR reflectivity profiles and model fits for the a-Si:H/Si samples. The graphs are offset from each other by a factor of 0.01 for clarity. Data and fits are not footprint corrected, but this affects only a part of the plateau of total reflection and is taken into account by the fitting program.}
    \label{xrays40}
\end{figure}

Eleven samples consisting of 40 nm a-Si:H deposited on crystalline Si substrates of size 3x3 cm$^2$ were measured with NR at the INTER time-of-flight reflectometer at ISIS, STFC Rutherford Appleton Laboratory, UK. Of these samples, ten had an a-Si:H layer were annealed at $T = 100$, $200$, $250$, $300$, $350$, $375$, $425$, $450$, $475$ and $500\,^{\circ}\mathrm{C}$, respectively. One sample was not annealed and served as a reference sample in the analysis. The reflectivity $R(q)$ was measured at two incidence angles of $\theta = 0.5\,^{\circ}$ and $2.3\,^{\circ}$, respectively, for a combined measurement time of about 1 hour and 20 minutes per sample. In addition, the samples were measured with a PANalytical X-ray reflectometer operating at the Cu $K_{\alpha}$ wavelength $\lambda = 1.542$ \AA. The specular reflectivity $R(\theta)$ was measured from $2 \theta = 0\,^{\circ}$ to $10\,^{\circ}$ for about 2 hours. NR and XRR data together with model fits of the reflectivity as a function of the transferred momentum $q = 4\pi {\lambda}^{-1} sin \theta $ are plotted in Figs. \ref{neutrons40} and \ref{xrays40}. Due to the large number of samples, data corresponding to four samples representative of the evolution by annealing temperature are shown in Fig. \ref{neutrons40} and the later Fig. \ref{sld40}.

\section{Results and Discussion}
As seen in Fig. \ref{neutrons40}, annealing the bilayer at $T = 300 \,^{\circ}\mathrm{C}$ does not change the NR results as compared to the reference sample, %with $T = 25 \,^{\circ}\mathrm{C}$
while annealing at $T = 450 \,^{\circ}\mathrm{C}$ smoothen out the diffraction fringes significantly which is further enhanced when annealing at $T = 500 \,^{\circ}\mathrm{C}$. In contrast, Fig. \ref{xrays40} shows that the XRR results are only modestly affected by the annealing. These observations are qualitatively consistent with the fact that hydrogen is desorbed at the higher temperatures, which reduces the neutron contrast between the a-Si:H layer and the substrate, but does not affect the X-ray contrast, which is caused by the lower density of the a-Si:H compared to the crystalline Si in the substrate.

%In order to quantitatively determine the H-distribution in the layers, the data was inverted, allowing the determination of the scattering length density as a function of the distance $z$ from the substrate.

NR and XRR data were simultanously fitted using the same four-layer model previously used in \cite{Qviller} %on a Si substrate of crystalline density $\rho = 2.330$ g/cm$^3$ in
using the GenX 2.4.9 reflectivity fitting package \cite{Bjorck}. The four layer model of the a-Si:H consists of a bottom layer composed of H and Si, then a H-depleted layer composed of H, Si and some O, capped with two oxide layers containing only SiO$_{2}$. For evaluating the goodness of fit, the logarithm of the reflectivity was used. This has the advantage of minimizing comparatively larger systematic errors at and around the critical edge. The density, thickness, and roughness of all layers were fitted using both the NR and XRR data. The coherent scattering length per formula unit was also fitted when analysing the neutron results. For the two bottom layers the densities were locked together during the fitting process.

\begin{figure}[!t]
  \centering
  \includegraphics[scale=0.37]{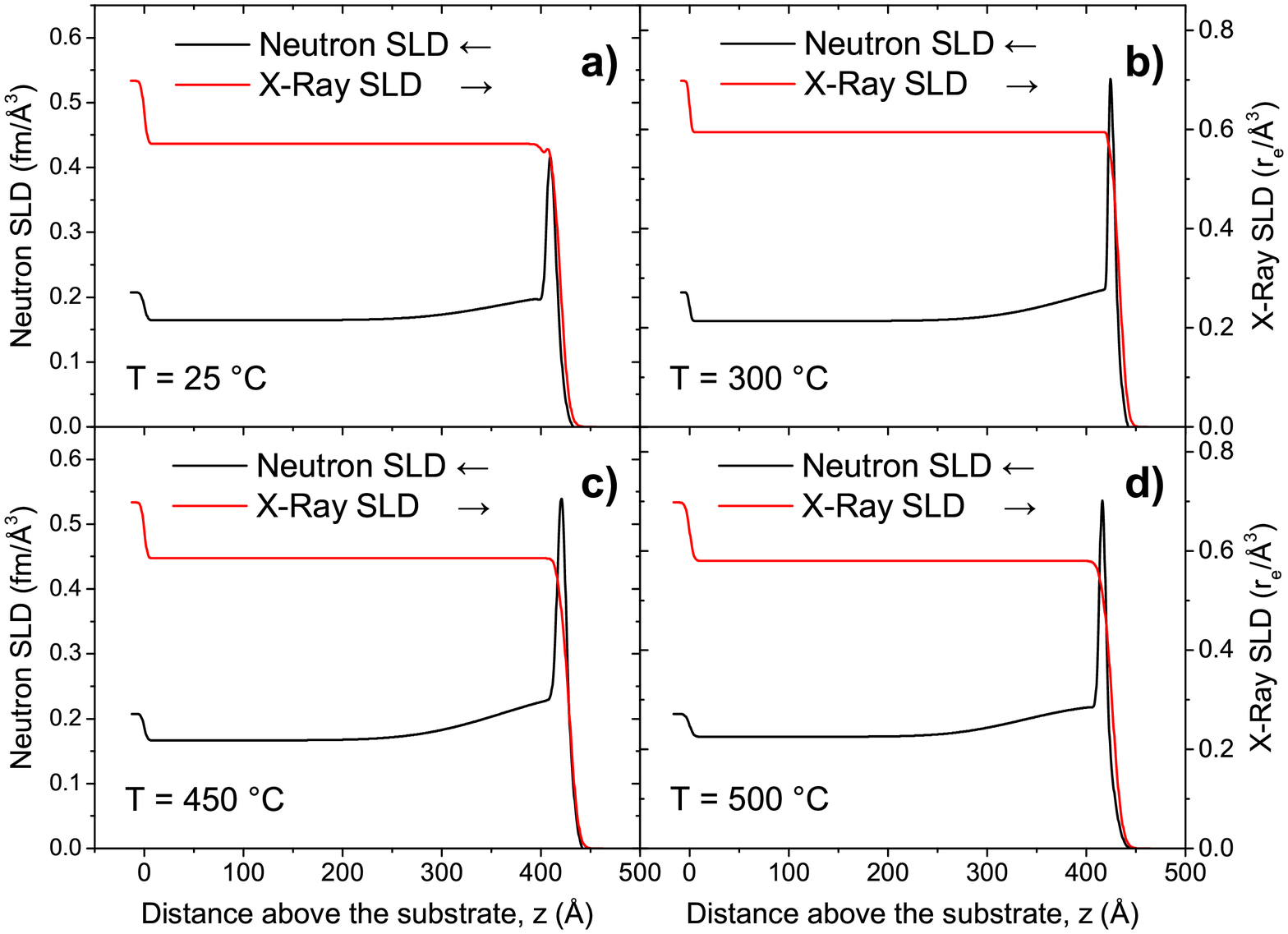}
  \caption{Neutron and X-ray SLDs for the a-Si:H/Si samples: a) $T = 25\,^{\circ}\mathrm{C}$, b) $T = 300\,^{\circ}\mathrm{C}$, c) $T = 450\,^{\circ}\mathrm{C}$ and d) $T = 500\,^{\circ}\mathrm{C}$.}
  \label{sld40}
\end{figure}

The resulting NR and XRR SLDs are plotted in Fig. \ref{sld40}. An innermost plateau is visible in the NR SLDs, corresponding to the bottom layer consisting of only Si and H. The NR data shows a large oxide peak, while the XRR SLDs are flat and then fall off due to the similarity of the scattering densities of Si and SiO$_{2}$. The oxide layers are probably not fully captured in the Parratt formalism, artefacts around the critical edge in both the NR and XRR data may be interpreted as resulting from cracks or non-Gaussian roughness. It was not possible to fit the NR data without allowing the oxide NR SLD assume to higher values for the coherent scattering length than stoichiometric SiO$_{2}$, resulting in a tall peak in the NR SLD better resolved in this work than \cite{Qviller} due to increased $q$-range in the NR data. These issues, particularly those at high $q$, corresponding to short distances, are not thought to significantly affect the determination of the H-concentration which is determined from the depth of the fringes at low $q$, and thus corresponding to the combined chemical- and density contrast between the crystalline Si substrate and the thick a-Si:H bottom layer consisting of only porous Si and H. There is not much qualitative change in the SLDs from $T = 25\,^{\circ}\mathrm{C}$ to $T = 500\,^{\circ}\mathrm{C}$, shown in Fig. \ref{sld40} (a-d).

H-desorption can however be detected in a quantitative treatment. From the definition of the SLD, Eq. \ref{SLD}, one can derive the following equation for the H/Si ratio $n_{H}/n_{Si}$ by a simple rearrangement, as in \cite{Qviller}

\begin{equation}
\frac{n_{H}}{n_{Si}}=\frac{SLD \times \phi - b_{Si}}{b_{H}}
\label{SLDrearr}
\end{equation}

\begin{figure}[!t]
  \centering
  \includegraphics[scale=0.35]{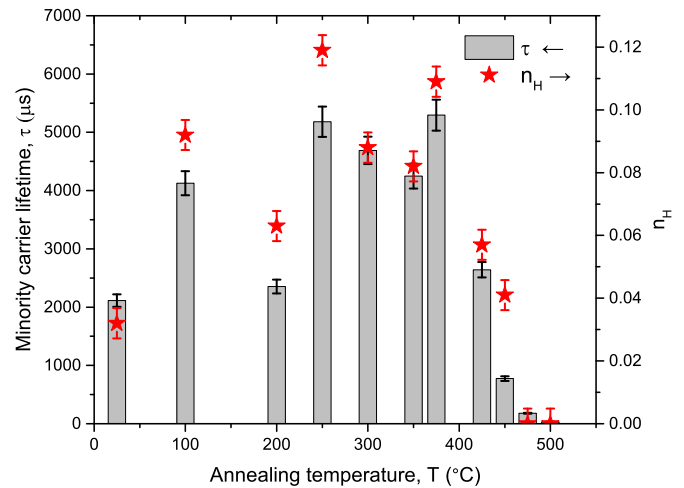}\\
  \caption{Minority carrier lifetime measurements $\tau$ for a-Si:H/Si samples annealed at different temperatures together with directly measured H-concentrations $n_{H}$ by reflectometry.}
  \label{lifetime40}
\end{figure}

By setting $n_{Si}$ = 1 and inserting $\phi$ and tabulated coherent scattering lengths $b_{Si} = 4.1491$ fm and $b_{H} = 3.7390$ fm, one can calculate the number of H atoms per Si atom from its determined SLD and density, $n_{H}$, in the innermost layer relevant for passivation. The molecular volume $\phi$ is futhermore defined as $\phi= M_{R}/ \rho N_{A}$, where $M_{R} = 28.09$ is the Si formula mass, $\rho$ is the Si density and $N_{A} = 6.022 \cdot 10^{23}$ is Avogadro's number. Furthermore, it has been assumed that H does not contribute to the molecular volume. The samples annealed at $T = 475$ and $500\,^{\circ}\mathrm{C}$ returned values of $n_{H}$ so small that they can be viewed to be zero within the uncertainties in the coherent scattering lengths and fitting procedure in general, thus $n_{H}$ was set to zero. Values of the fit parameters relevant for the calculation, the resulting $n_{H}$-values and minority carrier lifetimes $\tau$ measured by photoconductance are listed in Table \ref{values40} and plotted in Fig. \ref{lifetime40}. Error bars for the $n_{H}$ have been derived from noticing how much $\rho$ and the SLD can change before the fits get significantly worse, which is approximately 2\% for each parameter. As the covariance of the errors is not determined by a genetic algorithm, a 4\% error was applied to the largest $n_H$ value and this number was used as the error for all the $n_H$ values. Contrastingly, a 5\% relative error determined from experience was used for the $\tau$ values.

\begin{table}[]
\centering
\caption{Neutron SLD values, densities $\rho$, calculated H/Si ratio $n_{H}$ and minority carrier lifetimes $\tau$.}
\label{values40}
\begin{tabular}{|c|c|c|c|c|} \hline
$T(\,^{\circ}\mathrm{C})$ & SLD (fm/\AA$^3$) & $\rho$ (g/cm$^{3}$) & $n_{H}$ & $\tau$ ($\mu$s) \\ \hline

 25	& 0.165 & 1.91	& 0.032	             & 2113  \\ \hline
100	& 0.164 & 2.01	& 0.092	             & 4127  \\ \hline
200	& 0.167 & 1.99	& 0.063	             & 2353  \\ \hline
250	& 0.158 & 1.99	& 0.119	             & 5181  \\ \hline
300	& 0.163 & 1.99	& 0.088	             & 4689  \\ \hline
350	& 0.159 & 1.93	& 0.082	             & 4249  \\ \hline
375	& 0.162 & 2.02	& 0.109	             & 5295  \\ \hline
425	& 0.162 & 1.92	& 0.057	             & 2642  \\ \hline
450	& 0.167 & 1.95	& 0.041	             &  774  \\ \hline
475	& 0.167 & 1.87	& 0                  &  179  \\ \hline
500	& 0.172 & 1.94	& 0                  &   31  \\ \hline

\end{tabular}
\end{table}

\begin{figure}[!t]
  \centering
  \includegraphics[scale=0.35]{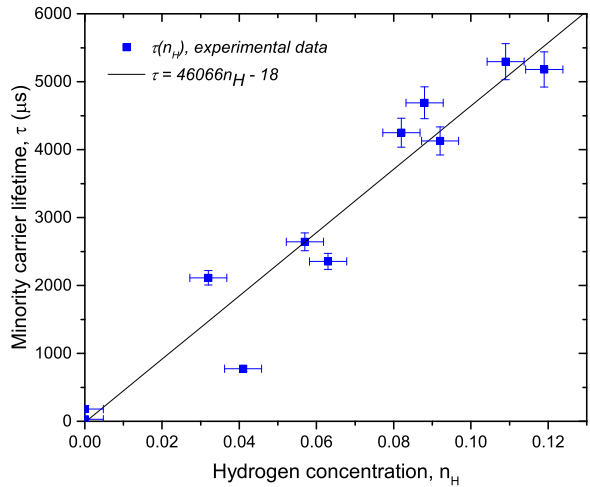}\\
  \caption{Correlation between H-concentration $n_{H}$ and minority carrier lifetime $\tau$ together with a linear fit $\tau = 46066n_{H}-18$.}
  \label{correlation}
\end{figure}

It is clearly seen in Fig. \ref{lifetime40} that $n_H$ and $\tau$ are strongly correlated and that the minority carrier lifetimes are dramatically reduced when H is desorbed at and above $T = 425\,^{\circ}\mathrm{C}$. Fig. \ref{lifetime40} show an evolution of $n_H$ and $\tau$ in temperature that resembles the evolution of $\tau$ in Fig. 1 in \cite{Qviller}. These considerations motivated an analysis of the correlation between $n_H$ and $\tau$, yielding a Pearson correlation $c = 0.96^{+0.03}_{-0.10}$ for the samples, where the errors on the correlation coefficient correspond to a 95 \% confidence interval. This implies that $92^{+6}_{-18}$\% percent of the variance in minority carrier lifetimes is explained by the H-content, the latter number obtained from the square of the correlation coefficient. In Fig. \ref{correlation} the minority carrier lifetime $\tau$ in microseconds ($\mu$s) is plotted as a function of the H/Si ratio $n_{H}$ together with a linear fit $\tau = 46066n_{H}-18$, where the tiny negative constant term is clearly unphysical and explained below. The linear relationship can be interpreted as one surface defect is passivated by a given number of hydrogen atoms, for example one to one ratio.

Apart from measurement errors and modelling assumptions, the small negative constant term in the linear fit can be traced down to the annealing at the medium investigated temperatures actually enhances the passivating effect of H to some degree. This effect is visible in the range from $T = 300$ to $400\,^{\circ}\mathrm{C}$ in Fig. 1 in \cite{Qviller} and at the data points at $T = 300$, $350$ and $375\,^{\circ}\mathrm{C}$ in Fig. \ref{lifetime40}. The three mentioned points in Fig. \ref{lifetime40} lie above the linear relation in Fig. \ref{correlation} and tilt the fit upwards. This results in a small, negative and unphysical constant term in the linear fit. A possible explanation of the minority carrier lifetime enhancement may be redistribution of hydrogen from oversaturated to undersaturated areas of the sample, which would in principle be detectable with off-specular reflectometry. Excluding these three temperatures with significantly enhanced passivation changes the linear fit to $\tau = 42459n_{H}+31$, removing the unphysical negative sign of the constant term. The correlation with uncertainties is now $c = 0.96^{+0.03}_{-0.16}$ and correspondingly, $92^{+6}_{-28}$\% of the minority carrier lifetime variation is explained by H-content.

\section{Conclusion}
In this work it has been shown with the use of NR, XRR and photoconductance measurements that a-Si:H is stable with regards to hydrogen desorption up to about $T = 425\,^{\circ}\mathrm{C}$. Furthermore, a strong correlation between H-concentration and minority carrier lifetimes has been established, allowing for a linear model fit and an interpretation in terms of a fixed number of H-atoms passivating one defect. Samples in the intermediate temperature regime are shown to have a slightly improved lifetime for the same hydrogen content due to an annealing effect. The developed procedure allows for quick, non-destructive measurements of H-concentration together with other material parameters of photovoltaic silicon that are otherwise hard to directly measure.

\section*{Acknowledgment}
The work was financed by The Research Council of Norway through the SYNKN\O YT program, Project No. 218418. Experiments at the ISIS Pulsed Neutron and Muon Source were
supported by a beamtime allocation (RB1510377) from the STFC (UK).

\ifCLASSOPTIONcaptionsoff
  \newpage
\fi

\end{document}